# Wavelets as q-bits and q-bit states as wavelets

Pawel Steblinski and Tomasz Blachowicz

Institute of Physics, Silesian University of Technology

Krzywoustego 2 str., 44-100 Gliwice, Poland

In this short report it is argued that by the use of wavelets formalism it is possible to describe the q-bit state. The wavelet formalism address the real-valued physical signals, for example, obtained during typical physical measurements.

## I. Introduction

Quantum computing and q-bit formalism bases on simple matrix operations. The q-bit is usually represented by a single-column or a single-row. A gate, which acts onto the q-bit state, can be written as a squared matrix. Thus, the quantum operations can be narrowed to multiplications and summations. The same fundamental operations are applied in digital wavelet formalism. Below, we try to describe q-bits adopting wavelet analysis of time series obtained in a typical physical experiment.

## II. Wavelets and qubits

Wavelet analysis is an approach which enables simultaneous analysis of signals in frequency and time domains, thus transients events can be interpreted from the frequency and time scales. Usually, the analysis is represented as a two-dimensional map which is constructed from an input signal. In Figs. 1a-1b the idea of wavelet analysis is shown. The two time-events, which amplitudes are well located at moment $t_1$ and $t_2$, respectively, have the two frequency values $\omega_1$ and $\omega_2$, and finally were localized on the time-frequency map as the $W_1$, $W_2$, and $W_3$ events. In other words, at the same $t_1$ moment, there are two physical events $W_1$ and $W_2$. On the other hand, the $\omega_1$ frequency is represented by the two events $W_1$ and $W_3$ separated in time (Fig. 1a). The spatio-temporal events can even have more subtle structure (Fig. 1b).

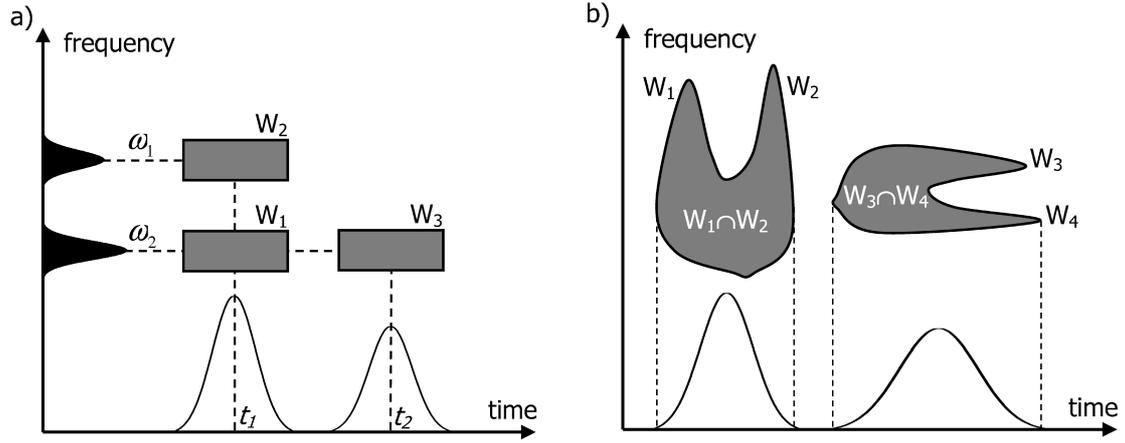

Fig. 1. The idea of 2-dimensional wavelet map. The two physical events $W_1$ and $W_2$ can be distinguished in the frequency scale, while the $\omega_2$ spectral component is composed from the two events, $W_1$ and $W_3$, separated in time moments $t_1$ and $t_2$ (a). The events can be or can't be separated in time and/or frequency (b).

In order to perform spectral wavelet analysis, the real-valued function is applied $g(z)$, which can be then convoluted with the physical signal $f(t)$. The result is represented by the wavelet spectral decomposition map $W(\omega,T)$ calculated as an integral in time, for the frequency $\omega$, at the given time-moment $T$.[1] Thus, we have

$$W(\omega,T) = \sqrt{\omega} \int_{-\infty}^{+\infty} f(t) g[\omega(t-T)] dt. \tag{1}$$

In order to extract from the above equation the physical signal $f(t)$ we need to calculate the inverse wavelet transform and obtain the dual function $\Psi[\omega(t-T)]$ which satisfies the following condition

$$\int_0^{+\infty} d\omega \int_{-\infty}^{+\infty} |\omega|^3 g[\omega(t'-T)] \Psi[\omega(t-T)] dT = \delta(t-t') \tag{2}$$

with the Dirac's delta function $\delta(t-t')$. Thus, we can express the original signal as the superposition of dual functions, namely

$$f(t) = \int_0^{+\infty} d\omega \int_{-\infty}^{+\infty} W(\omega,T)\, \omega \Psi[\omega(t-T)] dT. \qquad (3)$$

The above wavelet spectral decomposition can now be applied to the q-bit-like formalism. According to Ref. 2 the q-bit state can be defined as:

$$Q = \alpha \begin{bmatrix} 0 \\ 1 \end{bmatrix} + \beta \begin{bmatrix} 1 \\ 0 \end{bmatrix} = \alpha \vec{m} + \beta \vec{n}, \qquad (4)$$

where the $\alpha, \beta$ are complex numbers in general, and $\vec{m}, \vec{n}$ are the unit versors which span an elementary q-bit space.

Next, in order computing operations, we can set up the relation $R$ between the two $Q_1$ and $Q_2$ q-bits, considering entanglement or separation between them. It can written as follows

$$Q_1 R Q_2 = (\alpha_1 \vec{m}_1 + \beta_1 \vec{n}_1) R (\alpha_2 \vec{m}_2 + \beta_2 \vec{n}_2) = U_{11} \vec{n}_{11} + U_{12} \vec{n}_{12} + U_{21} \vec{n}_{21} + U_{22} \vec{n}_{22}, \qquad (5)$$

where

$$U_{ij} = U_{ij}(\alpha_i, \beta_i, \alpha_j, \beta_j) \qquad i,j = 1,2 \qquad (6)$$

are functions of complex numbers and where the unitary vectors $\vec{n}_{11}, \vec{n}_{12}, \vec{n}_{21}, \vec{n}_{22}$ represent in general the entanglement between q-bits. On the other hand, we can affirm that the two q-bits are separated if the above relation can be represented by the following simple expression

$$Q_1 R Q_2 = (\alpha_1 \vec{m}_1 + \beta_1 \vec{n}_1) \bullet (\alpha_2 \vec{m}_2 + \beta_2 \vec{n}_2), \qquad (7)$$

where the $\bullet$ symbol is the multiplication operator. We can state, however, that the two q-bits are entangled if they are, for example, the Bell states.[3] Eq. 5. represents the Bell states if it fulfills the following conditions

$$(U_{11} = 0 \wedge U_{21} = 0) \vee (U_{11} = 0 \wedge U_{22} = 0) \vee (U_{12} = 0 \wedge U_{21} = 0) \vee (U_{12} = 0 \wedge U_{22} = 0). \qquad (8)$$

Especially, from this equation we have

$$Q_1 R Q_2 = (\alpha_1 \vec{m}_1 + \beta_1 \vec{n}_1) R (\alpha_2 \vec{m}_2 + \beta_2 \vec{n}_2) = U_{12} \vec{n}_{12} + U_{22} \vec{n}_{22} \text{ or,} \qquad (9a)$$

$$Q_1 R Q_2 = (\alpha_1 \bar{m}_1 + \beta_1 \bar{n}_1) R (\alpha_2 \bar{m}_2 + \beta_2 \bar{n}_2) = U_{12} \bar{n}_{12} + U_{21} \bar{n}_{21} \text{ or,} \tag{9b}$$

$$Q_1 R Q_2 = (\alpha_1 \bar{m}_1 + \beta_1 \bar{n}_1) R (\alpha_2 \bar{m}_2 + \beta_2 \bar{n}_2) = U_{11} \bar{n}_{11} + U_{22} \bar{n}_{22} \text{ or,} \tag{9c}$$

$$Q_1 R Q_2 = (\alpha_1 \bar{m}_1 + \beta_1 \bar{n}_1) R (\alpha_2 \bar{m}_2 + \beta_2 \bar{n}_2) = U_{11} \bar{n}_{11} + U_{21} \bar{n}_{21} , \tag{9d}$$

what means finally that the entanglement can be still expressed in the same 2-dimensional space as for the single Q-bit state.

Adopting the wavelets decompositions, we can now define the q-bit (comp. Eq. 4) as

$$Q = W(\omega_i, T_j) \begin{bmatrix} 0 \\ \Psi(\omega_i, T_j, t) \end{bmatrix} + W(\omega_p, T_q) \begin{bmatrix} \Psi(\omega_p, T_q, t) \\ 0 \end{bmatrix} = W(\omega_i, T_j) \bar{m} + W(\omega_p, T_q) \bar{n} \tag{10}$$

where the meaning of versors $\bar{m}$ and $\bar{n}$ is obvious, namely

$$\bar{m} = \begin{bmatrix} 0 \\ \Psi(\omega_i, T_j, t) \end{bmatrix} \text{ and } \bar{n} = \begin{bmatrix} \Psi(\omega_p, T_q, t) \\ 0 \end{bmatrix}. \tag{11}$$

From the above we can see that the q-bit in general is a time-dependent object composed from the different physical frequencies, $\omega_i$ and $\omega_p$, connecting events from the different moment $T_j$ and $T_q$. Thus, the q-bit amplitudes are build from wavelet spectral decompositions $W(\omega, T)$, while the orthogonal state, which span the q-bit, are composed from the dual functions $\Psi(\omega, T, t)$.

### III. Conclusions

Within the current work the theory of wavelets involved in the q-bit states formalism was proposed. It was shown that the evolution of a q-bit structure, in time and frequency, can be considered as an application in quantum gates which can realize quantum computations. The stable states, on the other hand, can act as quantum memories. Our report has only preliminary character, however this topic will be developed in a near future.